\documentclass{article}

\usepackage[dvips]{color}
\usepackage{graphicx}

\begin{document}

\title{Selected Topics in Near Threshold Pion Photoproduction and Compton
Scattering off Nucleons\footnote{Summary of an invited talk presented at the 2nd MAX-LAB workshop on the nuclear physics program with real photons below 250 MeV; Lund May 30 - May 31, 2002. Preprint No. TUM-T39-02-17}}
\author{Thomas R. Hemmert \\
Theoretische Physik T39, Physik Department, TU M\"{u}nchen}

\maketitle

\begin{abstract}
Some open topics in the field of low energy photon-nucleon 
interactions are discussed---which in my opinion are of 
interest both for obtaining new information on nucleon structure and for precision tests of our theoretical understanding of chiral dynamics. In particular, I discuss p-wave multipoles in charged pion photoproduction off protons as well as the role of spin-dependent effects in (unpolarized) Compton scattering off nucleons.  There the concept of {\em dynamical} spin polarizabilities is found to be essential for the understanding of differential Compton cross sections above 120 MeV cms photon energy. 
\end{abstract}


\section{Introduction/Motivation}

Near threshold pion photoproduction as well as low energy Compton scattering
off single nucleons are certainly well established electro-nuclear processes
known already for several decades. The continuing interest in these
reactions at the beginning of the 21st century---from my personal
opinion---is twofold. On the one hand one wants to be sure to understand
these fundamental, ``simple'' interactions involving single nucleons very
precisely before one can draw any conclusions regarding the role of few-body
forces and/or (nuclear) many body effects when studying scattering processes
in more complicated systems like nuclei or nuclear ensembles like stars. The
other aspect has to do with our understanding of the nature of the strong
interaction. At this point in time QCD is the established theory for strong
interactions. However, its real world applications to issues typically
addressed in nuclear physics are rather limited, owing to its confinement
properties at low energies. In order to develop a systematic theoretical
description of the strong interaction in this crucial low energy domain,
already in the late 1970s pioneering studies were begun to map QCD onto a
field theory written in terms of its Goldstone Boson degrees of freedom.
Such an \emph{effective} field theoretic framework---usually\footnote{%
I prefer the term ``chiral effective field theory'' (ChEFT) instead of ChPT
to take into account the fact that during the past decade many
non-perturbative techniques have been applied to studies in ChEFT. The
unifying principles behind all these studies should be a general effective
Lagrangian framework consistent with the spontaneous + explicitly broken
chiral symmetry of QCD, written in terms of Goldstone Boson degrees of
freedom coupled to arbitrary external sources as well as matter fields. In
addition, a power-counting scheme needs to be specified, which enables the
theorist to select a finite number of contributing vertices from the
infinitely large most general effective chiral Lagrangian as well as allows
a systematic calculation of higher order corrections. Contrary to popular
opinion, a ``unique'' chiral power counting does not exist---at least not
for systems with more than 2 active flavors or more than one type of
external matter field. The label ``ChPT'' attached on a particular
calculation therefore always needs to be accompanied by the specification of
the power-counting employed.} in the literature called chiral perturbation
theory (ChPT)---was extended from the pure Goldstone Boson sector to more
complex scenarios involving one and more nucleons/baryons in the late 1980s,
respectively early 1990s. Nowadays, at the beginning of the 21st century, an
impressive variety of theoretical tools ranging from non-relativistic
perturbation theory to non-perturbative resummation methods are at the
chiral practitioners' disposal, covering physics topics from ultra low
energy hadronic atoms to nuclear matter systems above saturation density. At
the core of all these applications is the general chiral effective
Lagrangian of QCD, which connects all these seemingly disparate physical
scenarios via its chiral field tensors and coupling constants/counterterms.
Coming back to the issues discussed at this workshop in Lund, I went through
this introductory detour to emphasize that a high quality description in
terms of chiral effective field theory for such elementary reactions like
pion photoproduction or nucleon Compton scattering indeed has far-ranging
consequences for a much wider spectrum of physical processes\footnote{I 
consider
this feature to be the main difference between chiral effective field theory
and (chiral) modeling for these very processes.}. 
Admittedly, at this stage,
chiral effective field theory---despite its many stunning successes---is
certainly far from constituting a closed chapter in our understanding of the
strong interaction in its really interesting domain\footnote{%
For example, issues of different flavor and power-counting schemes, higher
order corrections, relativistic unitarization as well as new regularization
schemes certainly need to be further developed}. However, the prospect of
unifying large sections in the wide field of strong interaction
studies/nuclear physics into one systematic theoretical framework/lingua
franca constitutes a big attraction, that undoubtedly will lead to further
progress in theoretical understanding and technique. For this program to
proceed it is \emph{essential} that high precision low energy experiments
like pion photoproduction or nucleon Compton scattering continue to be
performed at accelerator laboratories like MAX-LAB. Only when the energy,
momentum, flavor and multipole dependence of the elementary reactions are
known to high precision chiral effective field theory can be further
developed and will thus improve its predictive power/theoretical soundness
in physically interesting scenarios where high precision data are much
harder or even impossible to come by. In the following I will focus on two
low energy electromagnetic interactions which are both interesting in their
own right and at the same time will also help to further constrain chiral
effective field theory---(charged) pion photoproduction as well as the role
of spin-dependent structures in nucleon Compton scattering.

\section{Near Threshold Pion Photoproduction}

\subsection{Multipole Truncation}

The general matrix element pertaining to pion photoproduction off a nucleon
contains 4 structure functions $F_{i}(E_{\pi },z)$. For a photon with
four-momentum $k^{\mu }=(\omega ,\vec{q})$ and polarization vector $\epsilon
^{\mu }=(\epsilon _{0}=0,\vec{\epsilon})$ producing a pion with four-momentum $%
q^{\mu }=(E_{\pi },\vec{q})$ in the cms frame it reads 
\begin{eqnarray}
M_{\gamma N\rightarrow N\pi } &=&\frac{M_{N}}{4\pi \sqrt{s}}\,T\cdot \epsilon
\nonumber \\
&=&\chi ^{\dagger }\{F_{1}(E_{\pi },z)\,\,i\,\vec{\sigma}\cdot \vec{\epsilon}%
+F_{2}(E_{\pi },z)\,\,\vec{\sigma}\cdot \hat{q}\,\,\vec{\sigma}\cdot (\hat{k}%
\times \vec{\epsilon})  \nonumber \\
&&+F_{3}(E_{\pi },z)\,\,i\,\vec{\sigma}\cdot \hat{k}\,\,\vec{\epsilon}\cdot 
\hat{q}+F_{4}(E_{\pi },z)\,\,i\,\vec{\sigma}\cdot \hat{q}\,\,\vec{\epsilon}%
\cdot \hat{q}\}\chi \;,
\end{eqnarray}
with $z=\cos \theta _{cms}=\hat{q}\cdot \hat{k}$. $\sigma ^{i}$ denotes the
Pauli matrix in spin space between the two-component spinors $\chi /\chi
^{\dagger }$of the incoming/outgoing nucleon.

Near threshold the expression given above can be simplified by performing a
multipole expansion and truncating the resulting amplitude at the p-wave
level. Performing the appropriate projections one finds 
\begin{eqnarray}
M_{\gamma N\rightarrow N\pi }^{l=1} &=&\chi ^{\dagger }[E_{0+}(E_{\pi
})\,\,i\,\vec{\sigma}\cdot \vec{\epsilon}+P_{1}(E_{\pi })\,\,i\,\vec{\sigma}%
\cdot \vec{\epsilon}\,\hat{k}\cdot \hat{q}+P_{2}(E_{\pi })\,\,i\,\vec{\sigma}%
\cdot \hat{k}\,\vec{\epsilon}\cdot \hat{q}  \nonumber \\
&&+P_{3}(E_{\pi })\,\,\vec{\epsilon}\cdot \left( \hat{q}\times \hat{k}%
\right) ]\chi  \label{sp}
\end{eqnarray}
Usually the predictions of chiral effective field theory calculations for
single pion photoproduction are then discussed at the level of the
contributing s-wave multipole $E_{0+}(E_{\pi })$, as well the 3 p-wave
multipoles\footnote{%
Historically the p-wave multipoles have been denoted by $E_{1+},\,M_{1+},%
\,M_{1-}$. They are connected to the structures $P_{i}$ introduced in Eq.(%
\ref{sp}) via 
\begin{eqnarray*}
P_{1}(E_{\pi }) &=&3\,E_{1+}(E_{\pi })+M_{1+}(E_{\pi })-M_{1-}(E_{\pi }) \\
P_{2}(E_{\pi }) &=&3\,E_{1+}(E_{\pi })-M_{1+}(E_{\pi })+M_{1-}(E_{\pi }) \\
P_{3}(E_{\pi }) &=&2\,M_{1+}(E_{\pi })+M_{1-}(E_{\pi })
\end{eqnarray*}
} $P_{i}(E_{\pi }),\;i=1,2,3$. One of the nice features of the p-wave
truncation of Eq.(\ref{sp}) is the resulting simple form for the \emph{%
unpolarized} differential cross section. To l=1 one obtains 
\begin{eqnarray}
\frac{d\sigma }{d\Omega }{\big |}_{l=1}=\frac{|\vec{q}|}{|\vec{k}|}\left[
A(E_{\pi })+B(E_{\pi })\,z+C(E_{\pi })\,z^{2}\right] ,
\end{eqnarray}
with 
\begin{eqnarray}
A(E_{\pi }) &=&|E_{0+}|^{2}+\frac{1}{2}\,|P_{2}|^{2}+\frac{1}{2}\,|P_{3}|^{2}
\nonumber \\
B(E_{\pi }) &=&2\,Re\left( E_{0+}P_{1}^{\ast }\right)  \nonumber \\
C(E_{\pi }) &=&|P_{1}|^{2}-\frac{1}{2}\,|P_{2}|^{2}-\frac{1}{2}%
\,|P_{3}|^{2}\;.
\end{eqnarray}
By fitting the three energy-dependent parameters $A,\,B,\,C$ to the
differential cross section one can therefore determine three bi-linear
combinations of the four (complex) low energy multipoles $%
E_{0+},\,P_{1},\,P_{2},\,P_{3}$. I will discuss the predictions of chiral
effective field theory for these quantities in the next two sections.

Finally, it is noted that important additional information pertaining to the
dynamics in the p-wave multipoles can be obtained by utilizing linearly
polarized photons measuring the photon asymmetry \cite{schmidt}. However, I
will not discuss this interesting possibility here as the plans for
polarized photons at MAX-LAB unfortunately will not cover the required
energy range in the near future \cite{LOIphotons}.

\subsection{Neutral Pion Photoproduction}

Within chiral effective field theory neutral pion photoproduction near
threshold is probably the best studied low energy electromagnetic production
process on a single nucleon. After all, it was in this reaction where it was
first shown that ChPT can also be utilized in the baryon sector to calculate
quark-mass dependent corrections beyond the venerable results of current
algebra in a systematic approach. On the theoretical side, both the s- and
the p-wave multipoles discussed in the previous section are now known to
next-to-leading one-loop order (i.e. $O(p^{4})$) within the SU(2) heavy
baryon approach \cite{BKM}. Further studies utilizing relativistic
frameworks and/or the inclusion of explicit Delta degrees of freedom are
underway, pointing again to the importance of this process both as a
benchmark for further development of theoretical tools as well as to its
importance as a basic building block in theoretical studies of
electromagnetic reactions in light nuclei. On the experimental side good
data in the threshold region now exist both for differential cross sections
as well as for the photon asymmetry, leading to an experimental determination
of all 4 s- and p-wave multipoles rather close (or extrapolated) to
threshold \cite{schmidt}. The agreement between theory and experiment is
impressive \cite{schmidt}.

What remains to be done in the neutral pion sector ? Aside from interesting
ideas to study the cusp behavior in detail \cite{bernstein}, I think it would
be interesting to have direct experimental information on the \emph{energy
dependence} of the $E_{0+},P_{i}$ multipoles or of the structure functions $%
F_{i}$. We note that (in isospin notation) in neutral pion photoproduction off
protons one is sensitive to the linear structure function combination 
\begin{equation}
F_{i}^{\gamma p\rightarrow p\pi ^{0}}=\left[ F_{i}^{(+)}+F_{i}^{(0)}\right]
,\;i=1,2,3,4\; ,  \label{proton}
\end{equation}
whereas for the (hypothetical) production process off a (free) neutron
target one finds 
\begin{equation}
F_{i}^{\gamma n\rightarrow n\pi ^{0}}=\left[ F_{i}^{(+)}-F_{i}^{(0)}\right]
,\;i=1,2,3,4\; .  \label{neutron}
\end{equation}
From my point of view, the really interesting part with respect to nucleon
structure is therefore encoded in the structure function $F_{i}^{(0)}$ (or
its $E_{0+}^{(0)},P_{i}^{(0)}$ components), as it is this piece that
dominates the difference between proton and neutron structure at low
energies (see also the discussion in section \ref{Compton}). While
theoretical analyses of neutral pion photoproduction off the deuteron with
the goal to gain insight into the production off the neutron are still
struggling with the proper separation of deuteron structure effects, another
way to improve our understanding of the makeup of the \underline{neutron}
could be provided by precision studies of
the energy dependence of the structure functions in neutral pion
photoproduction off \underline{protons} (Eq.(\ref{proton})). This suggestion 
is based on
the fact that the $(+)$- and the $(0)$-components actually have quite a
different energy dependence as well as different resonance components
associated with them \cite{BHM}. Given sufficiently accurate data from the
proton over a range of energies, one might be able to constrain the small 
$F_{i}^{(0)}$contributions
in this way and via Eq.(\ref{neutron}) thus also learn more about neutron
structure.

\subsection{Charged Pion Photoproduction}

In the case of charged pion photoproduction off hydrogen the four physical
structure functions $F_{i}(E_{\pi },z)$ for the process $\gamma
p\rightarrow n\pi ^{+}$ can be obtained from the isospin amplitudes via 
\[
F_{i}^{\gamma p\rightarrow n\pi ^{+}}=\sqrt{2}\left[ F_{i}^{(0)}+F_{i}^{(-)}%
\right] ,\;\;i=1,2,3,4 
\]
Given that here the $(-)$-components contribute to the process instead of
the $(+)$-components of the previous section, one should not be surprised
that---in addition to a different Born term structure---also \emph{the
one-loop contributions as well as the role of counter terms/short distant
physics are quite different in charged pion photoproduction compared to the
neutral case}. As far as sound predictions based on chiral effective field
theory are concerned, charged pion photoproduction (as far as its
non-trivial, non-Born contributions are concerned) is harder to deal with
than the neutral pion analogue. Already at the leading-one-loop (i.e. $%
O(p^{3})$) level in heavy baryon ChPT one has to deal with 5 different
counterterm contributions \cite{FHLU}. However, mainly based on data of the
inverse reaction $\pi ^{-}p\rightarrow \gamma n$ measured at TRIUMF,
predictions for the 4 s- and p-wave multipoles at threshold now exist, which
overall agree reasonably well with existing dispersion relation analyses for
these multipoles \cite{Lothar}, but also show interesting discrepancies by as
much as 30\%. Interestingly, these p-wave multipoles---to my
knowledge---have not yet been directly tested by experiment. Certainly the
s-wave multipole $E_{0+}$ at threshold has received repeated attention (it
is dominated by the Kroll-Ruderman term) and also is the topic of a
letter-of-intent at this workshop \cite{LOIpiplus}. However, in my opinion a
study of the p-wave multipoles---be it only very close 
threshold or maybe better
even with their energy-dependence---is much more interesting. A recent
analysis shows \cite{FHLU} that the relative importance of $O(p^{3})$
contributions---i.e. the interplay between counterterms and chiral
loops---in the case of charged pion photoproduction is larger in the p-wave
multipoles than in $E_{0+}$. I may also add---on a very speculative
note---that if an experiment really found discrepancies relative to the
currently accepted numbers (e.g. see \cite{Lothar}) for the s- and p-wave
multipoles of charged pion photoproduction (respectively their $(0)$- or $(-)$%
-components) in the near threshold region, such a find could also have
important consequences for nucleon (neutron ?) structure quantities like the
GDH-integral or the forward spin-polarizability, which are typically
evaluated via sum rules with the low energy components carrying a large
weight. Of course new experiments on charged pion photoproduction near
threshold would also provide important constraints for chiral effective
field theory, precisely \emph{because of} the delicate interplay between
short and long-distance physics in this reaction. I would therefore like to
ask my colleagues from the experimental side to give this reaction some
thought as to whether a good (even unpolarized) experiment can be done there.

\section{Low Energy Compton Scattering off Nucleons\label{Compton}}

\begin{figure}[!htb]
\begin{center}
\includegraphics*[width=12cm]{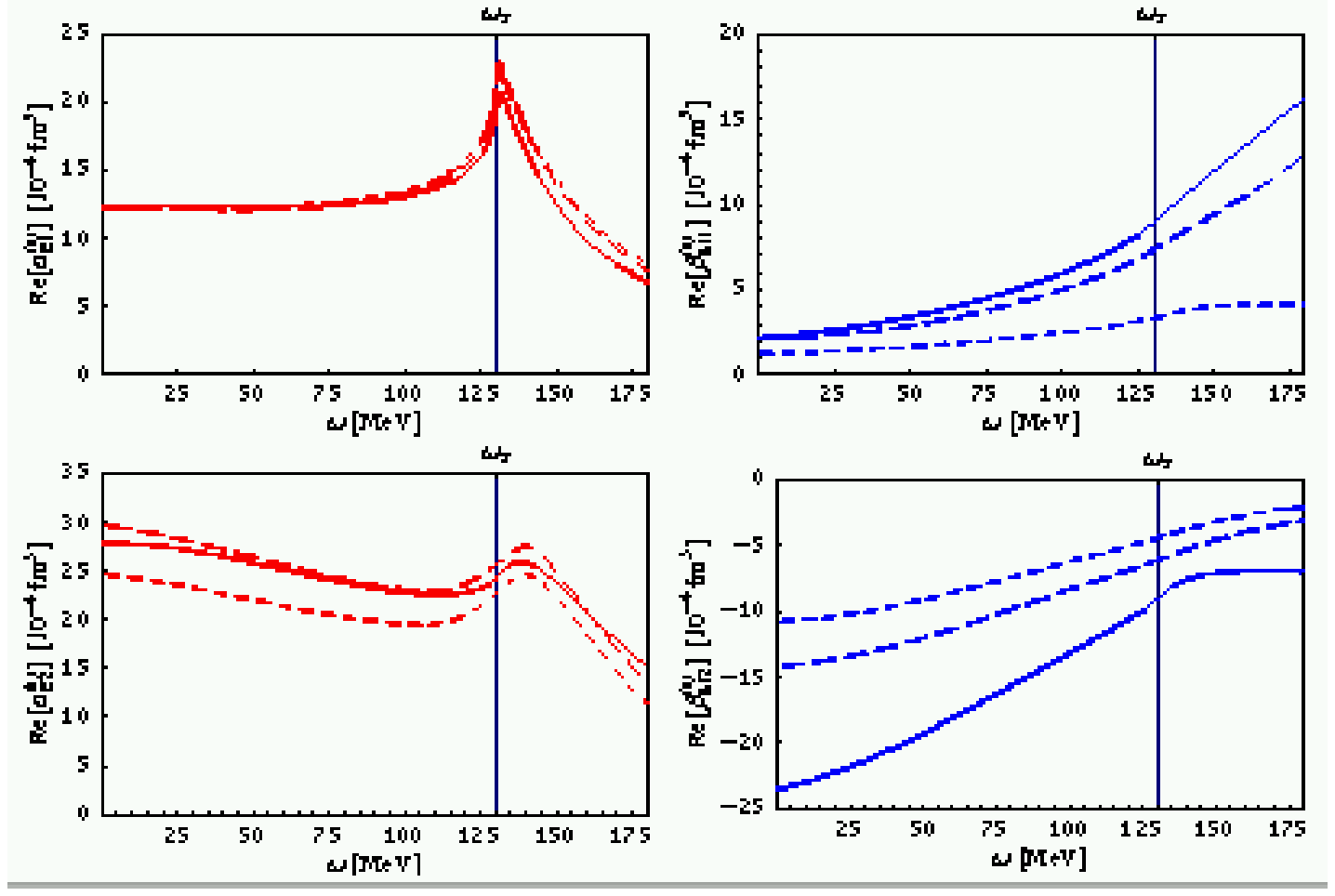}
\caption{Comparison between leading-one-loop SSE (long dashed curve), leading-one-loop heavy baryon ChPT (short dashed curve) and dispersion theory (full curve) for the isoscalar dynamical dipole and quadrupole polarizabilities of the nucleon.\cite{HGHP}}
\label{fig1}
\end{center}
\end{figure}

Beautiful data on Compton scattering off protons exist nowadays over a wide
range of energies and angles \cite{Olmos}. 
Unfortunately---for reasons that I have never
understood---most of the theoretical discussion in Compton scattering
focuses on just 2 numbers: The (static) electric and magnetic dipole
polarizabilities $\alpha _{E},\beta _{M}$. These static polarizabilities are
defined as the first structure dependent terms in a low energy expansion in
photon energy $\omega $ of the Compton cross section. However, this low
energy expansion itself (typically truncated at $\omega ^{2}$) is only valid
up to ca. 70 MeV photon energy, whereas most Compton data available hardly
start at such low energies---constituting a rather compromising mismatch
between a theoretical concept and experimental reality. In practice one has
dealt with this problem by utilizing the machinery of dispersion relations.
Using as input Compton data above pion threshold as well as available
information on single and double pion photoproduction, the dispersion
machinery has been the tool of choice to extrapolate Compton data from the
Delta region and beyond down to $\omega =0,$ where the static
polarizabilities are properly defined. Recently, it was pointed out \cite{GH}
that this energy dependence in its various multipole channels used for such
an extrapolation (but usually ``hidden'' in the dispersion machinery) is
interesting both in its own right and from the point of view of chiral
dynamics. It will not come as a surprise to professional physicists that the
task ``to get 2 numbers correct'' can be achieved within a large variety of
nucleon structure models, as such providing no definitive
insight into the physical
mechanisms at work in nucleon polarizabilities. In Ref.\cite{GH} we
therefore proposed the concept of ``dynamical polarizabilities'', based upon
the well-known multipole expansion for Compton scattering. These dynamical
polarizabilities are functions of the photon energy $\omega $ and, in the
limit $\omega \rightarrow 0$, reproduce the usual (static) polarizabilities.
The energy dependence of these polarizabilities is a measure of the
dispersion effects in that particular multipole channel, i.e. one expects to
see pertinent resonance features, relaxation phenomena or cusps associated
with the onset of inelasticities. For more details on the nature of
dispersion effects I refer to standard textbooks on classical
electrodynamics, the concepts taught there can all be carried over to a
discussion of the electromagnetic structure of the nucleon. Obviously, one can
also utilize 
dispersion theory to analyze the energy dependence of dynamical
polarizabilities---after all it had this very information built in from the
start. In Fig.\ref{fig1} I show a comparison \cite{HGHP} between
a dispersion analysis (solid curves), leading-one loop heavy baryon
ChPT (short dashed curves), as well as leading-one-loop SSE\footnote{%
SSE is a chiral effective field theory with definite powercounting that in
addition to the $\pi N$-dynamics of heavy baryon ChPT also employs explicit $%
\Delta $(1232) resonance degrees of freedom, for details see \cite{SSE}.}
 (long-dashed
curves) for the dynamical isoscalar\footnote{Note that chiral effective field 
theory predicts that isovector components in the (dynamical) polarizabilities 
of the nucleon are a higher order effect and therefore should be much smaller 
than their isoscalar counterparts. The dispersion analysis presented in 
\cite{HGHP} supports this finding, leading us to conclude that as for as the
 electromagnetic stiffness is concerned, proton and neutron should behave 
rather similar.} electric/magnetic dipole and quadrupole
polarizabilities of the nucleon. In the electric dipole polarizability all 3
approaches agree remarkably well, displaying an enormous cusp at the pion
production threshold. In the magnetic polarizability the SSE approach
and the dispersion analysis both predict a rapidly increasing paramagnetic
response of the nucleon when the photon energy goes up. This feature is due
to the strong $\gamma N\Delta $ M1 excitation, which in heavy baryon ChPT
only gradually gets built up via higher order terms. The dynamical electric
quadrupole polarizability $\alpha _{E2}$ also shows good agreement between
the dispersion analysis and SSE, while for the dynamical magnetic quadrupole
polarizability $\beta _{M2}$ all 3 frameworks differ. From the structures
displayed in Fig.\ref{fig1} it should be clear, that the very energy
dependence of these dynamical polarizabilities is largely due to chiral
dynamics ! With the concept of dynamical polarizabilities any theoretical
calculation of nucleon structure can now be put to a much more stringent
test. It%
\'{}%
s not enough anymore to reproduce roughly the right size for the static
values of the polarizabilities, one can now also ask the question whether
the mechanisms supposedly responsible for the electromagnetic stiffness of
the nucleon also reproduce correctly the associated energy 
dependence/dispersion effects.

\begin{figure}[!htb]
\begin{center}
\includegraphics*[width=12cm]{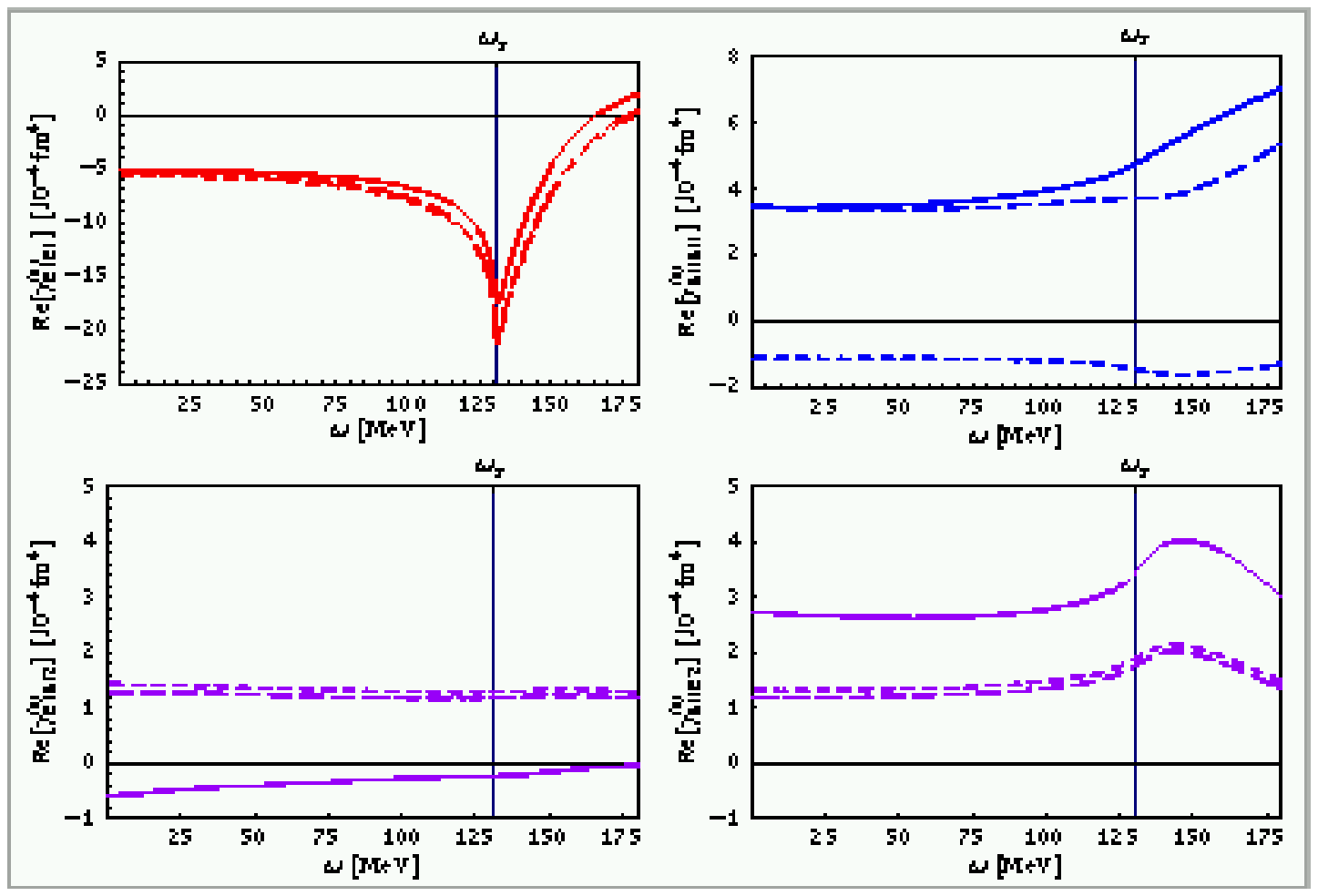}
\caption{Comparison between leading-one-loop SSE (long dashed curve), leading-one-loop heavy baryon ChPT (short dashed curve) and dispersion theory (full curve) for the isoscalar dynamical dipole spin-polarizabilities of the nucleon. \cite{HGHP}}
\label{fig2}
\end{center}
\end{figure}

How do we know what the correct energy dependence of the dynamical
polarizabilities looks like ? Amazingly, as shown in \cite{HGHP}, up to cms
photon energies of 170 MeV a truncation of the multipole expansion for
Compton scattering at l=1 is entirely consistent with all existing
(unpolarized) Compton data on the proton ! Remember, the dynamical
polarizabilities are defined as functions of energy. This finding therefore
offers the possibility to choose one particular cms photon energy (e.g. $%
\omega _{0}=150$ MeV), replace the theoretical predictions at this
particular energy by 6 free parameters\footnote{%
While 6 parameters might be too many to be determined just from the angular
dependence at one particular energy $\omega _{0}$ , it would already
constitute a big step forward if one can constrain several linear
combinations of these parameters in an unpolarized Compton experiment. A
complete separation of all 6 dipole structures presumably can only be
obtained if one also resorts to polarized Compton scattering. Studies in
this direction are under way \cite{polarized}.} and try to determine these 6
parameters from the angular dependence measured at that energy. In such a
fashion one could attempt to experimentally map out the energy dependence of
the dynamical dipole polarizabilities at different points $\omega _{0}$ and
in this way test what we as theorists are dreaming up. 
Note that we have 6 structures
to deal with because at l=1, in addition to the dynamical electric and
magnetic dipole polarizabilities $\alpha _{E1}(\omega _{0}),\beta
_{M1}(\omega _{0}),$ one also has to take into account the 4 dynamical
dipole spin-polarizabilities\footnote{%
One can show \cite{HGHP} that in the limit $\omega \rightarrow 0$ one is
able reproduce the static dipole spin polarizabilities obtained in the
leading-one-loop SSE calculation of Ref.\cite{HHKK}.}: $\gamma
_{M1M1}(\omega _{0}),\gamma _{E1E1}(\omega _{0}),\gamma _{M1E2}(\omega
_{0}),\gamma _{E1M2}(\omega _{0})$. A comparison between predictions from
dispersion analysis, leading-one-loop heavy baryon ChPT and leading-one-loop
SSE for these 4 elusive nucleon spin structure components is shown in Fig.%
\ref{fig2} \cite{HGHP}. We note that for the purely electric and the purely
magnetic dipole spin-polarizabilities SSE and the dispersion analysis are
again in good agreement, while chiral effective field theory and the
dispersion analysis systematically differ for the mixed dipole
spin-polarizabilities. At the moment it is not clear whether this deviation
is entirely due\footnote{%
Certainly in $\gamma _{M1E2}$ an excitation/deexcitation of the Delta
resonance via a combination of M1 and E2 transitions is not yet taken into
account at leading-one-loop order in either of the 2 chiral effective field
theories considered here, as the $\gamma N\Delta $ E2 transition only comes
in at higher orders in the chiral expansion.} to missing higher order terms
in the chiral effective field theories---experimental input regarding these
essentially unknown nucleon structure quantities would therefore be
extremely interesting.

\begin{figure}[!htb]
\begin{center}
\includegraphics*[width=12cm]{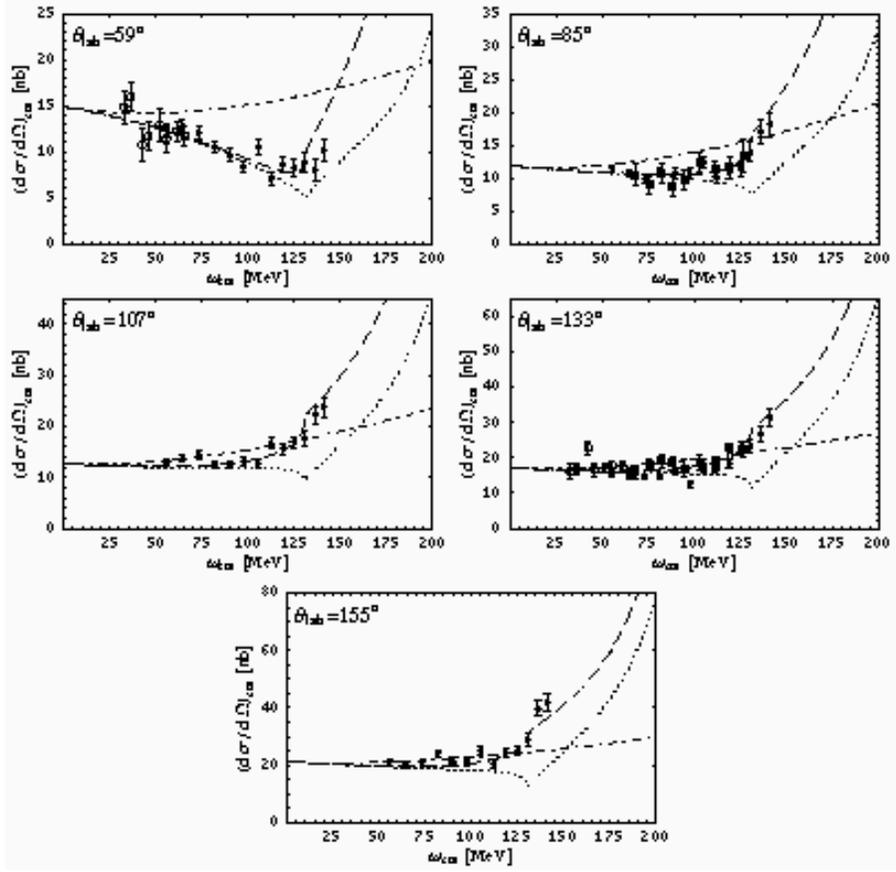}
\caption{Dash-dotted curve: l=0 (Born term) result for the Compton cross section. The complete l=1 leading-one-loop SSE prediction is shown as the dashed curve, whereas in the dotted curve all spin-polarizabilities have been set to zero (by hand). Surprisingly, there seems to exist good sensitivity to the (dynamical) spin-polarizabilities for photon enrgies above 120 MeV (cms).\cite{Robert}}
\label{fig3}
\end{center}
\end{figure}

Finally, I want to close this presentation with a brief discussion of Fig.%
\ref{fig3} \cite{Robert}. The dash-dotted curve shows the theoretical
prediction for the Compton differential cross-sections at l=0, i.e. all
polarizabilities are set to zero, only the Born terms are kept. The
interesting feature in this plot can be seen from the difference between the
dotted and the dashed curves. The dashed curve constitutes the full l=1
leading-one-loop SSE prediction (which, as already mentioned, though being
truncated at l=1 provides an excellent description of the data), whereas the
dotted curve also comes from l=1 leading-one-loop SSE, but with the four
dynamical dipole spin-polarizabilities discussed above by hand forced to be
zero. We therefore conclude that even in an unpolarized Compton experiment,
assuming that one knows the dynamical spin-independent dipole
polarizabilities  $\alpha _{E1}(\omega ),\beta _{M1}(\omega )$ reasonably
well, there seems to be interesting sensitivity\footnote{%
Note that the difference between the 2 curves in Fig.\ref{fig3} is not due
to large effects associated with the pion pole, as this structure is
subsumed in our Born terms. (This is also the reason why the l=0
truncation---i.e. all polarizabilities set to zero---provides such a decent
description of the data in the backward direction.} to the dynamical dipole
spin polarizabilities in the photon energy region above 120 MeV (cms).
Clearly, a detailed analysis together with experts from the experimental
side has to be performed to properly evaluate this exciting possibility.

In summary, I want to emphasize that in past few decades of Compton
scattering off the proton we have heard a lot about electric and magnetic
polarizabilities of the nucleon. However, the spin polarizabilities are
still largely uncharted territory and present a real frontier for the field
of nucleon structure studied with electromagnetic probes. I hope that the
field takes up the challenge and attempts to measure these structures, even
if this requires polarized Compton experiments. I am also optimistic that
some of the puzzling results in Compton scattering off the deuteron \cite
{Bent} will appear in a new light, once the here present leading-one-loop
SSE framework  is combined with a chiral potential ansatz. Work along these
lines is beginning this fall \cite{Dan}. 

\bigskip

\bigskip

\begin{center}
\textbf{Acknowledgments}
\end{center}

\bigskip

I want to thank the organizers for inviting me to this stimulating
workshop at MAX-LAB. The work presented here was supported by BMBF
and DFG. I am grateful to ECT* for its hospitality when this summary was 
written up. I also want to thank N. Kaiser for a careful reading of this 
manuscript.

\newpage

\end{document}